\documentclass[12pt,a4paper]{article}

\title{\bf Comment on an alleged refutation of non-locality }
\author{R. L. Schafir\footnote{e-mail: roger.schafir@lgu.ac.uk}
\\\normalsize\textit{CISM, London Guildhall University, London EC3N 1JY, U.K.}\date{}}
\begin{document}
\maketitle
\begin{abstract}
\noindent A recent claim (Deutsch and Hayden (2000)), that
non-locality can be refuted by considering the evolution of the
system in the Heisenberg picture, is denied.  What they
demonstrated was not the falsity of non-locality but the
no-superluminal-signalling principle.
\end{abstract}
\hfil\vskip\baselineskip
\par
    Deutsch and Hayden (2000) have claimed that Bell's Theorem (or its later variants;
    for a relatively recent account, see Peres 1993), namely that non-locality follows from an entanglement situation, is a mistake.  They argue that if the evolution of the system into an entangled state is viewed in the Heisenberg picture rather than the Schrödinger picture, it can be seen that all effects are transferred sub-luminally.
\par
The purpose of this Comment is to point out that the result they
have actually obtained is not that non-locality is false, but the
standard theorem that no superluminal signalling is possible.  We
should remember that in addition to non-locality there is this
other result about entangled systems, which, remarkably, co-exists
with it, and that is the fact that, so long as each observer
simply looks at his or her own results, irrespective of the
results of the other observer, then the results are indeed
completely local, and in the example of EPR-Bohm are simply a
half-and-half random mixture of the two possible outcomes.
Non-locality, by contrast, is all in the correlations and has a
retrospective character: after the measurements have been made and
sufficient time has passed for the results of the observers to be
brought together and compared, then it is deduced that a non-local
effect \textit{must have} occurred.
\par
Deutsch and Hayden's article is expressed in the language of
quantum computing, with the entangling evolution into a Bell basis
state (Braunstein, Mann and Revzen, 1992) obtained as a
composition of the evolution caused by two types of quantum logic
gate.  However a first point to be made is that non-locality
follows solely from the correlations between the results of later
disentangling measurements; it does not matter how the pairs came
to be entangled in the first place, whether by an evolution or a
projection, and for that reason alone considerations of the
Heisenberg versus the Schrödinger picture would seem to be
irrelevant.
\par
But let us follow the authors in considering the particles to be
in a constant Heisenberg state, taken to be in the form before
they became entangled, and the spin observables to evolve through
the entangling evolution.  We will still, of course, obtain the
same probabilities when it comes to the disentangling
measurements, taking the mod-squared inner product of the constant
state with the (inversely) evolved spin-eigenstates, instead of
the evolved state with the constant spin eigenstates.  Following
the usual proof of non-locality, we wish to find the probabilities
of the outcomes of both observers being the same and of them being
different, when they measure at angles $\theta$ and $\phi$
respectively. So, since the individual eigenvalues are $\pm1$, we
evolve the product observable whose $\pm1$ eigenstates represent
'same' and 'different', namely
\begin {equation}
\left(\widehat{\sigma}_{z}\cos\theta+\widehat{\sigma}_{y}\sin\theta\right)\otimes\left(\widehat{\sigma}_{z}\cos\phi+\widehat{\sigma}_{y}\sin\phi\right)
\end{equation}
where the $\widehat{\sigma}$'s are the spin observables.  (Or we
could evolve all tensor combinations of the $\widehat{\sigma}$'s
and then take a combination appropriate to rotations, if
preferred.)

\par
But Deutsch and Hayden's expression for the evolved observables is
\begin{eqnarray}
\widehat{1}\otimes\left(\widehat{\sigma}_{x}\otimes\widehat{1},
-\widehat{\sigma}_{y}\otimes\widehat{1},
-\widehat{\sigma}_{z}\otimes\widehat{\sigma}_{x}\right)\otimes\widehat{1}
\nonumber\\\widehat{1}\otimes\left(\widehat{\sigma}_{x}\otimes\widehat{\sigma}_{z},
-\widehat{\sigma}_{y}\otimes\widehat{1},
-\widehat{1}\otimes\widehat{\sigma}_{x}\right)\otimes\widehat{1}
\end{eqnarray}
(their expression (23)), which is obtained by evolving the
observables
\begin{equation}
\widehat{\sigma}_{x}\otimes\widehat{1},
\widehat{\sigma}_{y}\otimes\widehat{1},
\widehat{\sigma}_{z}\otimes\widehat{1},
\widehat{1}\otimes\widehat{\sigma}_{x},
\widehat{1}\otimes\widehat{\sigma}_{y},
\widehat{1}\otimes\widehat{\sigma}_{z}
\end{equation}
(with rotations then applied).  But these give the \textit{total}
probabilities for \textit{each} individual observer's results
\textit{irrespective} of the results of the other observer.  These
are indeed local, but that is precisely the
no-superluminal-signalling theorem.
\par
Note that each product observable in (3) contains the spin
observable for one observer, and a sum over the projectors for the
two possible results of the other observer, represented by the
identity operator for the other observer.  In the same way, when
Deutsch and Hayden demonstrate the independence of the
probabilities for one observer from the choice of angle of the
other observer, by evaluating the expectation value for the one
observer (their Expression (26)), this summation does not appear
explicitly, but it is there, because in evaluating
\begin{equation}
<0|<0|U^{†}\left(\widehat{\sigma}\otimes\widehat{1}\right)U|0>|0>
\end{equation}
(where U is the evolution) the identity $\widehat{1}$ is again a
summation over the projectors for the second observer.  This is
precisely how the no-superluminal-signalling theorem is proved, by
taking the results of one observer and summing over the results
for the other observer (Shimony 1984).
\par The authors go on to
claim that teleportation, as well as non-locality, can be seen to
be a purely local phenomenon as a result of their analysis.
However their further analysis depends on a technique which seems
strange. They define recursively a set of evolutions $U_{G}(t)$
associated with logic gate G, for discrete t's, by\newline\\
$U_{G}(t+1)$ = same functional form in terms of updated
$\widehat{\sigma}$'s (i.e. of the Heisenberg-evolved observables
of the system) as $U_{G}(t)$ was in terms of previously updated
$\widehat{\sigma}$'s.\newline
\par
But why should we regard the updated mapping UG as representing
the same logic gate G?  A logic gate is presumably going to be a
fixed device which is always going to have the same effect on
bits, not a different effect at different times.  This puts an
obstacle, for myself at least, to following their argument on
"locally inaccessible information."
\par
But in any case, teleportation, although another entanglement
effect, is very different to non-locality, and it is not clear
that it requires explanation in the same way that non-locality
does.  Non-locality is an effect at the macroscopic level of
observed outcomes and observer choices of what to measure; it
follows purely from the correlations, and although quantum
mechanics predicts those correlations, non-locality would hold for
any other theory which predicted the same correlations, or would
hold even if those correlations had been established only by
experiment, without any idea of the mechanism between them.  Also,
it requires that the observers should be able to make free choices
of what to measure, since it consists of the assertion that for at
least one of the observers (though we don't know which one, and it
may be both), the observer's outcomes \textit{would have been}
different (note the contrafactual reasoning) if the other observer
had made a different choice of what to measure.  If the choices
were somehow pre-determined, for instance, there would be no
problem of non-locality.
\par
Teleportation, by contrast, is an assertion about copying a
quantum state, something which is not itself observable (indeed,
part of the idea is that the original state may be unknown, and
remains unknown after being copied), and the observers have no
choice of what to measure.  This gives no clear ground for
asserting that a superluminal influence crosses space at the stage
when the entanglement is "swapped" (to use the terminology of Pan
\textit{et al} (1998)), and in addition there must then be a
sub-luminal message before the copying is accomplished.
\par
It is non-locality which is definitely a problem, and since it
follows from correlations at the macroscopic level, any denying of
it requires either that the correlations are incorrect, and hence
that any theory which predicts them (in particular quantum
mechanics) is incorrect\footnote {And that the experiments which
have apparently confirmed them are invalid through loopholes}, or
denying the validity of the reasoning which establishes
non-locality as a consequence of the correlations, such as denying
the validity of contrafactual reasoning.

\end{document}